\documentclass[preprint,12pt]{elsarticle}

\usepackage{amssymb}

\usepackage{lineno}
\modulolinenumbers[5]

\usepackage{booktabs} 
\usepackage{array} 
\usepackage{colortbl} 
\usepackage{xcolor} 
\usepackage{multirow} 
\usepackage{amsmath} 
\usepackage{float} 
\usepackage{url}

\usepackage{graphicx}
\usepackage{rotating}
\usepackage{tikz}
\usepackage{placeins} 

\begin{document}

\begin{frontmatter}



\title{Forensic-Ready Risk Management Concepts}


\author[inst1]{Lukas Daubner}
\ead{daubner@mail.muni.cz}
\author[inst1]{Martin Macak}
\ead{macak@mail.muni.cz}
\author[inst2]{Raimundas Matulevi\v{c}ius}
\ead{raimundas.matulevicius@ut.ee}
\author[inst1]{Barbora Buhnova}
\ead{buhnova@mail.muni.cz}
\author[inst1]{Sofija Maksovi\'{c}}
\ead{maksovic@mail.muni.cz}
\author[inst1]{Tomas Pitner}
\ead{tomp@mail.muni.cz}

\affiliation[inst1]{organization={Faculty of Informatics, Masaryk University},
             addressline={Botanicka 68a}, 
             city={Brno},
             postcode={60200},
             country={Czechia}}
            
\affiliation[inst2]{organization={Institute of Computer Science, University of Tartu},
            addressline={Narva mnt 18}, 
            city={Tartu},
            postcode={51009},
            country={Estonia}}

\begin{abstract}
Currently, numerous approaches exist supporting the implementation of forensic readiness and, indirectly, forensic-ready software systems. However, the terminology used in the approaches and their focus tends to vary. To facilitate the design of forensic-ready software systems, the clarity of the underlying concepts needs to be established so that their requirements can be unambiguously formulated and assessed. This is especially important when considering forensic readiness as an add-on to information security. In this paper, the concepts relevant to forensic readiness are derived and aligned based on six existing approaches. The results then serve as a stepping stone for enhancing Information Systems Security Risk Management (ISSRM) with forensic readiness.
\end{abstract}




\end{frontmatter}

\section{Introduction}

Forensic-ready software systems are a category of systems that includes the capabilities to conduct sound digital forensic processes and generate sound digital evidence~\cite{Pasquale:2018}. To elicit and assess the specific requirements for the concrete software systems, a risk-based approach was suggested~\cite{Daubner:2021}. Currently, there are numerous implementation guidelines focusing on a general notion of forensic readiness available. However, the approaches tend to work with different concepts and different terminology.

One of the key challenges in systematically approaching forensic-ready software systems is identifying and aligning the fundamental concepts. The importance is even more relevant as the aim is to extend Information Systems Security Risk Management (ISSRM)~\cite{Dubois:2010,Mayer:2009} with the notion of forensic readiness, seeking to address threats in a robust risk-based approach. Therefore, the resulting method needs to precisely fit into the scope of security risk management, correctly reflect the concepts of forensic readiness and address the gaps concerning forensic-ready software systems.

This paper contains the first fundamental step of the methodology, where the relevant approaches for planning and implementing forensic readiness are processed. The aim is to identify and align the concepts relevant to a risk-based approach and extension of ISSRM for forensic-ready software systems. This methodology is based on the one by~\cite{Dubois:2010,Mayer:2009}, which was utilised to formulate the original ISSRM.

Understanding the underlining concepts requires extensive exploration and analysis of the forensic readiness approaches. Moreover, as the terminology is not established, the context for each concept is needed for its alignment. Most importantly, utilising the systematic methodology is essential to provide strong validation and generalisability of the resulting approach.

\section{Forensic Readiness Guidelines}

The state-of-the-art implementation guidelines for forensic readiness lean more toward establishing organisational processes. However, they are acknowledging the need for both technical and organisational controls for effective implementation.

\paragraph{Tan} Originally, forensic readiness is approached as a set of measures for network design and operations by Tan~\cite{Tan:2001}. The main focus is put on logs and logging mechanisms within a computer network and hosts, especially concerning Intrusion Detection Systems (IDS). The paper discusses more concrete technical controls and good practices to use in achieving forensic readiness rather than general concepts. However, several concepts essential for risk-based planning are mentioned or are generalisable. Particularly emphasised is the implementation of centralised, multi-tiered logging to support the validation of the primary data from multiple points. Furthermore, the timely establishment of highly trustable (potential) evidence is recommended using digital signatures and trusted timestamping. While no process to achieve forensic readiness is discussed, there are hints on identifying evidence sources and their content, typically logs.

\paragraph{Rowlingson} Evolving from the base notion, Rowlingson~\cite{Rowlingson:2004} introduces a ten-step implementation guide to achieve forensic readiness. The guide is aimed at enterprises and organisations with a mature security-minded culture. While recognising the importance of technical controls in the implementation of forensic readiness, they are viewed as a secondary measure. Primarily, the focus is put on organisational and procedural controls. Most importantly, however, risk assessment is considered the core activity from which the implementation plans can be derived. It further assumes that security controls, based on security risk assessment, are sufficiently implemented. The guide serves as a basis for a state-of-the-art approach for forensic readiness, frequently expanded upon (e.g.,~\cite{Sachowski:2019}).

In fact, the first step towards forensic readiness involves risk assessment, where the risks stem from crimes and disputes impacting business processes and their continuity. Specifically, the risk assessment is used to identify business scenarios where there is a benefit of having digital evidence. The listed benefits relevant from a security perspective are proving and reducing the impact of computer-related crime. However, the benefits go beyond security, including handling disputes, demonstrating compliance and handling data release court orders.

Following the identification of scenarios, the guide prescribes an enumeration of available sources of potential evidence by computer systems. Furthermore, detailed information about the sources is mandated, including location, format, retention, owner, and presence of personal information. Then, it must be decided if and how are the evidence sources helping in the context of the identified scenarios to formulate a requirement on evidence collection. Such requirement is evaluated based on the costs and benefits the collection would pose. Furthermore, factors influencing the utility and reliability of the potential evidence are recommended to consider.

With the formulated requirements, concrete capabilities for gathering digital evidence should be established. Again, there is an emphasis on the non-interference of business processes. Furthermore, the gathered evidence must be admissible, which mandates the employment of further security measures for evidence authenticity, integrity, and redundancy. Such measures must be kept during transfer and in long-term storage so that the admissibility of evidence is not degraded. To that end, the guide recommends the establishment of policies.

The subsequent steps of the guide discuss the implementation of supporting organisational policies and procedures. Concretely, alignment with security monitoring, escalation policies, staff training, evidence presentation, and legal review. While directly tied to forensic-ready software systems, these procedures are critical for successfully utilising the implemented capabilities.

\paragraph{Grobler et al.} Establishing a forensic readiness capability in an organisation is a goal of a general framework by Grobler et al.~\cite{Grobler:2010}. The focus of the framework is to formulate and align organisational controls as a proactive measure before an incident or evidence release order. It includes the steps to prepare an infrastructure for digital forensic investigation, an evidence management plan, the inclusion of evidentiary requirements into risk management, and allocation of personal capacity together with their training. Furthermore, it formulates the points of alignment between forensic readiness objectives and the information technology and security objectives.

While the framework does not specify the exact process for forensic-ready risk management, the identification of evidence and scenarios, or technical controls, it provides a common ground for their employment in a wider scope. From the perspective of forensic-ready software systems, the key step is the actual implementation of evidence requirements and an evidence management plan formulated based on the framework.

\paragraph{Elyas et al.} Another general forensic readiness framework formulated by Elyas et al.~\cite{Elyas:2015} also focuses on the organisational perspective. The framework aims to define a common, high-level vocabulary for concepts playing a role in organisational forensic readiness. These concepts are represented in a hierarchy with highlighted relationships on each level.

Primarily, the framework encapsulates the reasons why forensic readiness should be implemented by the organisation in terms of forensic readiness objectives. These are meant to guide the further steps of the implementation, namely the establishment of forensic readiness capability. However, it is noted that the established capability can influence the objective like a feedback loop.

The second part of the framework, forensic readiness, capability encapsulates the ability and means of the organisation to meet the objectives. It is further divided into organisational factors and forensic strategy. The organisational factors represent the management of the forensic readiness program and culture, which influences the forensic strategy. Notably, the authors note possible feedback from the implemented forensic strategy on organisational factors.

Finally, the strategy itself encapsulates the implementable aspects of forensic readiness and the effect on forensic and non-forensic stakeholders. It must be designed in accordance with the set forensic readiness objectives, and it is noted as unique for each organisation. The implementable concepts in the strategy are policy, training, and infrastructure, which is composed of technology and architecture. Concerning the forensic-ready software systems, the forensic infrastructure has the greatest overlap in scope. Specifically, forensic technology contains the employment of forensic tools, logging, and event reporting, while forensic architecture focuses on the overall environment and its capability to produce and handle potential evidence.

Again, the framework does not specify how the forensic strategy, specifically forensic infrastructure, should be implemented. However, utilisation of risk management is recommended to identify concrete cases.

\paragraph{Good Practice Guide 18} An extensive Good Practice Guide 18~\cite{GPG18:2015} published by the United Kingdom’s National Technical Authority for Information Assurance provides guidance in deploying forensic readiness policy and supporting practice. It is supplemented by Implementation Guide 18~\cite{IG18:2015}, which provides more concrete advice on the planning and implementing of forensic readiness. However, by the time of writing this paper, both guides are archived with no up-to-date replacement.

The guide establishes common concepts and vocabulary, business drivers, and principles of implementing forensic readiness. However, the focus is on organisational policy, with little attention given to systems. Essentially, forensic readiness is understood in terms of contingency planning and preparation of crisis scenarios in the case of indecent. The guide contains a quick overview of the digital forensics field and an elicitation of benefits of forensic readiness, risks if there is no such capability, as well as costs of its implementation. Most importantly, the guide defines twelve principles of forensic readiness policy, which create a rough planning and implementation process.

The state of forensic readiness within the organisation is expressed by capability level. A specific desired level is chosen as a part of the implementation process. For each capability level, there is a set of forensic readiness generic requirements that should be refined.

Continuing in the process, the core components of planning are the investigation scenarios. They describe a situation or an incident requiring an investigation that is relevant to the target organisation. Each scenario contains typical detection indicators, evidence sources, investigation workflow, and desired outcomes. As such, they are mostly focused on planning incident response and investigation. Furthermore, the scenarios are used for training purposes. Generic examples of the scenarios and investigation techniques are presented in the supplementary guide.

While not discussing the implementation of controls in software systems, the guide provides a solid foundation for the identification of specific forensic-ready requirements. Importantly, the definition and utilisation of scenarios can be expanded by mapping them onto the software systems. Forensic-ready requirements can be derived based on such mapping to meet desired outcomes of the scenario more reliably.

\paragraph{ISO/IEC 27043:2015, 30121:2015} Forensic readiness is discussed as a part of international standard ISO/IEC 27043:2015~\cite{ISO27043:2015} Incident investigation principles and processes. It presents the common digital forensic investigation processes at an abstract level, defines basic principles, and provides guidelines for incident investigation. Four plus one processes are defined on the highest level, each subsequently decomposed in more detail. Namely, the classes are readiness for pre-incident processes, initialization for processes in the initial stages of an investigation, acquisitive for potential evidence identification and handling processes, and investigative for processes uncovering the potential digital evidence. The additional class consists of concurrent processes cross-cutting through the main four classes. Furthermore, the standard aims to harmonise various investigation process models into a comprehensive model.

From the perspective of forensic-ready software systems, the class of readiness processes is the most relevant. However, the standard explicitly states that readiness class is optional, as the investigation can commence without preparation. The process class is further divided into three process groups focused on planning, implementation, and assessment in this order. Furthermore, the readiness processes are defined as iterative, meaning that the assessment process group feeds back to any of the previous processes if unsatisfactory.

After the identification and planning steps, their outputs are implement\-ed. Specifically, the standard prescribes implementation of the designed architecture, which consists of the deployment of controls supporting the subsequent implementation processes. Those entail the implementation of the proactive collection, handling, and storage of potential evidence, including logging controls and evidence protection controls. Furthermore, the pre-incident potential evidence analysis and incident detection controls are implemented according to the plans.

Finally, within the assessment process group, the satisfiability of the implementation is evaluated. Primarily, the implemented controls should be validated with respect to the goals. Furthermore, the conformity with the legal environment and regulations should be assessed as well.

There is another ISO/IEC standard focusing solely on forensic readiness, ISO/IEC 30121:2015~\cite{ISO30121:2015} Governance of digital forensic risk framework. It is explicitly focused on establishing a framework for forensic readiness, utilising risk management. However, the standard does not go beyond a brief definition of terms, principles, steps, processes, and metrics. It focuses more on establishing common basic terminology and governance principles for senior management. Concreate approaches on how to conduct risk management for forensic readiness are not discussed.

Both standards are defined in a broader spectrum of standards covering various aspects of digital forensics. Most importantly, ISO/IEC 27037~\cite{ISO27037:2012} concerning first response and initial evidence capture, ISO/IEC 27041~\cite{ISO27041:2015} concerning assurance of digital forensic methods and tools, and ISO/IEC 27042~\cite{ISO27042:2015} concerning the implementation of digital forensic methods and tools for evaluation, interpretation, and reporting of evidence.

In summary, the ISO/IEC 27043:2015 discusses the planning and implementation of forensic-ready software systems. Additionally, it recommends the utilization of risk management to identify and prioritize the scenarios requiring digital evidence and concrete controls. However, it does not goes into much detail on the exact plausible method.

\section{Alignment between Forensic Readiness and Security}

To justify the inclusion of forensic readiness into ISSRM, the relationship between forensic readiness and security is briefly discussed. Primarily, the focus is on the concerns of secure and forensic-ready software systems.

Forensic readiness is commonly approached jointly with security, more precisely as its enhancement. ISO/IEC 27043:2015~\cite{ISO27043:2015} states preservation or improvement of information security in the organisation systems as one of the aims of forensic readiness. There exist numerous overlaps in concerns between the two fields, and as a result, forensic readiness was proposed as a component of security best practice~\cite{Grobler:2007}. For example, auditing and monitoring controls are employed for incident detection and as important sources of potential evidence in forensic-ready software systems. Likewise, protection of potential evidence integrity can be achieved by security controls, with only a different reason for their employment.

However, the overlaps also cause conflicts. The most notable is conflict in post-incident recovery~\cite{Rowlingson:2004}. While forensics is concerned with sound preservation and thorough investigation, incident handling focuses on a fast recovery, which might spoil the evidence. Nonetheless, this advocates for joint consideration to find an acceptable trade-off.

Due to the discussed influences, security and forensic readiness are considered jointly to address the forensic-ready software systems. This is especially important in the context of insider attacks~\cite{Rowlingson:2004}, as pure security controls are often insufficient. For example, legitimate access might be exploited for an attack, or security policy might not be neglected. On the other hand, forensic readiness can allow for reliable detection and investigation of such cases. A similar proposal was made to cover the residual security risks utilising forensic-ready controls~\cite{Daubner:2021}.

\section{Forensic Readiness Concept Alignment}

\begin{sidewaystable}[htbp]
  \caption{Alignments of relevant concepts for Forensic-Ready Risk Management}
  \label{tab:conceptAlignments}
  \centering
  \scriptsize
  \begin{minipage}{\linewidth}
  \begin{center}
  \begin{tabular}{
  |>{\centering\arraybackslash} m{0.10\linewidth}
  |>{\centering\arraybackslash} m{0.12\linewidth}
  |>{\centering\arraybackslash} m{0.12\linewidth}
  |>{\centering\arraybackslash} m{0.12\linewidth}
  |>{\centering\arraybackslash} m{0.12\linewidth}
  |>{\centering\arraybackslash} m{0.12\linewidth}
  |>{\centering\arraybackslash} m{0.12\linewidth}|}
  \hline
  Concept & Tan~\cite{Tan:2001} & Rowlingson~\cite{Rowlingson:2004} & Grobler et al.~\cite{Grobler:2010} & Elyas et al.~\cite{Elyas:2015} & GPG 18~\cite{GPG18:2015} & ISO/IEC~\cite{ISO27043:2015,ISO30121:2015} \\
  \hline
  Potential Evidence & Incident Data, Log & Potential Evidence & Evidence & Potential Digital Evidence & Digital Evidence & Potential Digital Evidence \\
  \hline
  Evidence Source & Source of Evidence & Source of Potential Evidence & X & X & Digital Evidence Source & Potential Source of Digital Evidence \\
  \hline
  Evidence Storage & Storage for Log Data & Secure Storage & X & X & X & X \\
  \hline
  Event & Incident & Incident, Criminal Act & X & X & X & Incident \\
  \hline
  Impact & X & Impact & X & X & Impact, Harm & X \\
  \hline
  Risk & X & Risk & Risk & X & Risk & Risk \\
  \hline
  FR Scenario & X & Business Scenario Benefiting from Evidence & Incident & X & Scenario & Scenario \\
  \hline
  FR Goal & X & Benefit & X & Forensic Readiness Objective & Benefit, Scenario Class & X \\
  \hline
  FR Treatment & X & X & X & X & X & X \\
  \hline
  FR Requirement & X & Evidence Collection Requirement & Evidence Requirement & X & X & X \\
  \hline
  FR Control & X & Evidence Gathering Capability & X & Forensic Technology~/ Architecture & X & Control \\
  \hline
  \end{tabular}
  \end{center}
  \footnotesize
  \quad\quad X - Means there is no corresponding concept
  \end{minipage}
\end{sidewaystable}

By analysing the approaches for implementing forensic readiness, various common concepts were identified. The aim is to align those concepts in order to approach the forensic-ready systems in a systematic manner that allows formulating requirements to tackle the specific threats at hand. In this regard, the goal is to extend the ISSRM approach with forensic readiness. For a summary, the concept alignment is present in Table~\ref{tab:conceptAlignments}.

The concepts of Risk, Event, and Impact were taken directly from the ISSRM, as those are risk management concepts. Therefore, the established ISSRM definitions could be explored in their usage in the forensic readiness context without overriding their meaning in security. Additionally, the idea of risk treatment-related concepts from ISSRM is adapted to look for, not necessarily overlapping, equivalents in forensic readiness.

Naturally, the concept of Potential Evidence is present in every approach. Although they typically vary in emphasis on adjectives for “potential” or “digital” evidence. However, Tan uses the term incident data for the same context. Two other concepts that are semantically close to potential evidence and also tend to differ in adjectives are the Evidence Source and Evidence Storage.

Forensic-Readiness Scenario is identified as a key concept which typically drives the process. However, while very similar, the approaches adopt different notions, which must be aligned. Rowlingson~\cite{Rowlingson:2004} defines the scenarios as follows:

\begin{quote}
  \small
  ``What is the threat to the business and what parts are vulnerable? This is, in effect, a risk assessment, and is performed at the business level. The aim is to understand the \textbf{business scenarios} where digital evidence may be required and may benefit the organisation in terms of:...''~\cite{Rowlingson:2004}
\end{quote}

The scenarios are considered in connection with \textit{risk} and \textit{potential evidence}. However, the risks are approached from a business perspective rather than security, focusing on the \textit{benefits} of implementing forensic readiness. The need for potential evidence is then evaluated with respect to the scenario.

On the other hand, Grobler et al.~\cite{Grobler:2010} focus primarily on the security risks. Following this notion, the authors use the term \textit{incident} to describe the corresponding concept. It is connected to the \textit{potential evidence} (labelled as evidence or comprehensive digital evidence) and a \textit{risk}. The incident is commented on as follows:

\begin{quote}
  \small
  ``The authors propose to link evidence and process requirements for an identified risk to the risk profile of an \textbf{incident}. Apply an algorithm to calculate the completeness and admissibility of the evidence for the incident.''~\cite{Grobler:2010}
\end{quote}

GPG18~\cite{GPG18:2015} adopts a more abstract notion of scenario, where there could be a need for forensic investigation, therefore \textit{digital evidence}. A recommended approach is to build them based on identified \textit{risks}. Although the guide lists the \textit{benefits} of forensic readiness, the scenarios are not explicitly connected to them. Instead, they are divided into \textit{classes} based on what they involve (e.g., crime, civil dispute). In IG18~\cite{IG18:2015}, the scenario is defined as follows:

\begin{quote}
  \small
  ``\textbf{Scenarios} are defined as a compact summary of events with sufficient detail to make them effective for planning purposes. They should also be optimised, without over-sophistication, as the objective is to construct a limited series of scenarios in which an organisation may become involved, in order to reveal the common threads between investigations of that type.''~\cite{IG18:2015}
\end{quote}

Similarly, ISO/IEC 27043:2015~\cite{ISO27043:2015} also considers scenarios that might require \textit{digital evidence}. They are used as a component in the risk assessment to derive concrete controls. While the relationship between scenario and \textit{risk} is not explicitly stated, the connection to assets is. It is mentioned as follows:

\begin{quote}
  \small
  ``(In this process), one should examine all probable \textbf{scenarios}, where digital evidence might be required. (...) It is also recommended that a proper risk assessment is performed during this process for each identified scenario respectively. A risk assessment would enable one to better identify possible threats, vulnerabilities and related scenarios that would expose particular information assets.''~\cite{ISO27043:2015}
\end{quote}

It is further observed that the forensic-ready scenario is often formulated based on an abstract Forensic-Readiness Goal, describing the reason why forensic readiness is considered. As discussed, Rowlingson~\cite{Rowlingson:2004} considers the benefits of having digital evidence for a scenario. GPG18~\cite{GPG18:2015} also considers similar benefits, but it provides distinct classification directly for scenarios. A framework by Elyas et al.~\cite{Elyas:2015}, while not considering any concept comparable to a forensic-ready scenario, defines abstract forensic readiness objectives.

Lastly, akin to risk treatment-related concepts, Forensic-Ready Requirement and Control are identified. Although the term “requirement” was present almost in every approach, only some represent a step toward the implementation, as meant by ISSRM. A similar situation is with the controls, as the approaches often do not deal with the technical implementation of the plans. Curiously, Tan~\cite{Tan:2001} is explicit only in the evidence gathering requirement and controls.

Numerous gaps are found in the approaches. Firstly, they typically focus on the planning and organisational perspective rather than the technical side of forensic-ready software systems. The most overarching ones considering the technical controls are Rowlingson~\cite{Rowlingson:2004} and ISO/IEC 27043:2015~\cite{ISO27043:2015}. Secondly, there is a lack of clearly defined concepts. Thirdly, while risk management is often an integral part of the approaches, there is a discrepancy in whether to approach it from a security or business perspective and their alignment. And lastly, the approaches do not establish ways to evaluate the implemented controls.

\section{Conclusion}

In this paper, relevant forensic readiness concepts from various forensic readiness guidelines were identified, discussed, and aligned. Notably, the alignment was made in the broader risk management context. The purpose of this alignment is twofold. First, it helps to organise the terminology of forensic readiness. Second, it is an important stepping stone for a systematic approach to forensic-ready software system design. In this regard, the concept alignments allow for an extension of the Information Systems Security Risk Management (ISSRM) with forensic readiness concepts.


\section*{Acknowledgment}

\noindent This research was supported by ERDF "CyberSecurity, CyberCrime and Critical Information Infrastructures Center of Excellence" (No. CZ.02.1.01/\allowbreak0.0/0.0/16\_019/0000822).

%

\bibliographystyle{elsarticle-num}
\bibliography{bibliography}





\end{document}